\documentclass[12pt]{article} 

\begin{document}
\setcounter{page}{0}

\thispagestyle{empty}
\begin{flushright}
                                                   IITK-HEP-99-53  \\
                                                   hep-ph/9908205 \\
\end{flushright} 
\begin{center}
{\LARGE\bf  
$\mathbf{\rho}$ Parameter Constraints on Models with Large Compact
Dimensions}
\vskip 25pT
{\large\sl  Prasanta Das~\footnote{E-mail: pdas@iitk.ac.in} {\rm and}
Sreerup Raychaudhuri~\footnote{E-mail: sreerup@iitk.ac.in} }
\vskip 5pT
{\rm
Department of Physics, Indian Institute of Technology, \\ Kanpur 208
016, India.} \\
\vskip 10pt
{\large\bf Abstract} 
\end{center} 
\vskip 5pt
\begin{quotation} {\rm 
In models with large extra dimensions, where quantum gravity effects
become strong at the TeV scale, the $\rho$ parameter can receive large
contributions from one-loop diagrams involving exchange of multiple
graviton and dilaton states. These contributions are computed, taking into
account cancellation of spurious infrared divergences, and the (finite)
results for $d = 5$ and 6 are compared with current experimental data. It
is shown that 5 large extra dimensions are incompatible with the data and
$d = 6$ is severely constrained.  }
\end{quotation}
\vskip 120pt 
\begin{flushleft}
April 2000
\end{flushleft} 
\vfill

\newpage 

\begin{center}
{\large\bf 1. Introduction} 
\end{center}

Considerable excitement has been generated by the suggestion~\cite{ADD1}
that strong gravitational effects could become manifest at scales of the
order of a few TeV. One of the ways in which this can happen requires
large compactified spacetime dimensions beyond the four Minkowski
dimensions. An elegant model involves a string theory~\cite{GrScWi}
(living in 10 dimensions) which has solitonic excitations of the
gravitational field called $D$-branes~\cite{Polchinski}.  In this model,
proposed~\cite{ADD2} by Antoniadis, Arkani-Hamed, Dimopoulos and Dvali
(AADD), all Standard Model (SM) fields are conceived of as living on a
heavy $D_3$-brane embedded in a space of 10 dimensions. In fact, the SM
fields correspond to excitations of open strings whose ends are confined
to the brane. Of the extra 6 dimensions, it is possible to have $d$ of
them compactified with radii $R_c$, where $R_c$ can be large, while the
remaining (6-$d$) dimensions are compactified with radii near the Planck
length $M_{Pl}^{-1} \sim 10^{-33}$ cm. The latter do not play much part in
the subsequent discussion, though their existence is essential to have a
consistent string theory in the first place.  For all practical purposes,
therefore, spacetime consists of ($4+d$) dimensions, the extra (spatial)
$d$ dimensions being compactified, typically on a torus $T^{(d)}$ with
radius $R_c$ each way. This is called the {\em bulk}, as opposed to the
{\em brane} on which the observable Universe lives.  Since gravity
experiments have not really probed the sub-millimetre

The actual value of Newton's constant $G_N^{(4+d)}$ in the bulk can be
taken as large as ${\cal O}(1~{\rm TeV}^{-2})$, but its value $G_N^{(4)}$
in the effective 4-dimensional space at length scales $\gg R_c$ is the
extremely small one ${\cal O}(10^{-32})$ TeV$^{-2}$ measured in gravity
experiments.  The two are related, using Gauss' Law~\cite{ADD3}, by 
$$
\bigg[ M_{Pl}^{(4)} \bigg]^2 \sim R_c^d \bigg[ M_{Pl}^{(4+d)} \bigg]^{2+d}
$$ 
where $M_{Pl}^{(4+d)} \simeq \bigg[ G_N^{(4+d)} \bigg]^{-\frac{1}{2+d}}$ 
denotes the Planck mass
in the relevant number of dimensions.  If $M_{Pl}^{(4+d)} \sim 1$ TeV,
then $R_c \sim 10^{30/d - 19}$ m, {\it i.e.}, for $d = 1$, $R_c \sim
10^{11}$ m, which predicts deviations from Einstein gravity at solar
system scales. Since no such effects are seen, we are
constrained\footnote{Of course, for $d = 1$, we have $R_c \propto
M_S^{-3}$: thus $M_S > 10^4$ TeV would make $d = 1$ viable. However, then
there are no interesting collider signals.} to take $d \geq 2$.  For these
values $R_c < 1$ mm and there is no conflict with known
facts\cite{LoChPr}.  The smallness of Newton's constant $G^{(4)}_N$ is thus 
a direct consequence of the compactification and hence there is no hierarchy
problem in this theory\footnote{A related problem, that of stabilization
of the compactification scale, exists, however; this has been discussed in
Ref.~\cite{ArDiMaRu}. In fact, this is the chief criticism  of the AADD
model.}.

In traditional Kaluza-Klein (KK) theories~\cite{KK}, the mass-spectrum of
non-zero KK modes is driven to the Planck scale $M_{Pl}^{(4)}$. This
problem is avoided in the AADD model by having the SM particles live on a
`wall' with negligible width (which we identify with the $D_3$~brane).
While SM fields are confined to the brane, gravitons correspond to
excitations of closed strings propagating in the bulk. The only new
effects observable on the brane --- to which corresponds the observable
Universe --- will be those due to exchange of gravitons between SM
particles on the brane.

To construct an effective theory in 4 dimensions, {\em i.e.} on the brane,
gravity may be quantized taking the usual weak-field (`linearized') limit,
assuming that the underlying string theory will ultimately take care of
the well-known ultraviolet problems.  Interactions of gravitons now follow
from the ($4+d$)-dimensional Einstein equations in the compactification
limit. Feynman rules to the lowest order in $\kappa = \sqrt{16\pi G^{(4)}_N}$
for this effective theory have been worked out in detail in
Refs.~\cite{GiRaWe} and \cite{HaLyZh}. We make use of the prescriptions of
Han, Lykken and Zhang~\cite{HaLyZh} in our work. For convenience, 
some of the relevant Feynman rules are listed in Appendix A.

In this effective theory, the couplings of the gravitons to the SM
particles will be suppressed by the Planck scale $M^{(4)}_{Pl} \simeq 1.2
\times 10^{19}$ GeV. This is offset, however, by the fact that, after
compactification, the density of massive KK graviton states 
is very high, being indeed, proportional to
$\left[M^{(4)}_{Pl}\right]^2$. The $M^{(4)}_{Pl}$ dependence cancels out,
therefore, leaving an interaction of electroweak strength, whose scale is
set by the bulk Planck's constant $M_S \sim M^{(4+d)}_{Pl}
\sim 1$ TeV, henceforth called the `string' scale.  A further
assumption usually made is that the brane itself is heavy and its
vibration modes decouple from the processes under consideration. This
corresponds to a static approximation for the brane~\cite{Sundrum}.

The interactions generated by gravity between SM particles on the brane
can be written mostly 
in terms of exchange of spin-0 and spin-2 gauge bosons. At
the lowest order in $\kappa$, the spin-1 gauge bosons
decouple~\cite{ADD3,Sundrum} from matter due to the diagonal structure of
the energy-momentum tensor for matter fields on the brane.  Interactions
of the spin-0 dilaton (or `radion', $R$) and the spin-2 gravitons($G$) --- 
or rather, their massive KK modes --- have been described in
Refs.~\cite{GiRaWe} and \cite{HaLyZh} and several phenomenological studies
{\em at the tree-level}~have been made using these interactions. The
salient features of such studies are briefly surveyed in the next section.
However, it suffices to note that almost all of these tend to provide {\em
lower} bounds on the string scale $M_S$.

One-loop effects of the interactions of towers of KK states of the
graviton and radion have not yet been investigated in detail. The
pioneering work in this direction was in section 3.5 of
Ref.~\cite{HaLyZh}, in which a leading-order calculation of self-energy
corrections to the mass of a scalar particle were worked out. A more
detailed calculation of the anomalous magnetic moment of the muon was
attempted in Ref.~\cite{Graesser}. In the latter, the result turned out to
be `remarkably finite', though the actual numbers were rather
disappointingly small. It was argued in Ref.~\cite{Graesser} that one-loop
effects are worth calculating even though there may be other
TeV-suppressed operators at the tree-level, simply because the one-loop
effects are completely calculable, given the Feynman rules. One can add
the argument that a study of one-loop effects brings out subtle features of
the theory in a way that tree-level effects can never be expected to do.

In the present work, therefore, we compute one-loop corrections to the
masses of the electroweak gauge bosons $W$ and $Z$, which are constrained
by the famous $\rho$ parameter~\cite{rho}. Extra contributions to the
$\rho$ parameter have been discussed in the context of AADD-type models in
the literature~\cite{precision}, but not in the context of one-loop
calculations. It turns out that the one-loop contributions from virtual
radion and graviton states are strongly divergent in the ultraviolet
(UV), a feature which is to be expected in an effective theory. The use of
the string scale $M_S$ as a momentum cutoff has been suggested in
Ref.~\cite{HaLyZh} (and elsewhere) and this leads to large contributions
which actually {\em grow} with increasing $M_S$.  Recognising that this is
not a pathological feature but simply a manifestation of the fact that we
have an effective (non-renormalisable) theory, we can therefore, use the
experimental bounds on the $\rho$ parameter to derive upper bounds on this
scale $M_S$, unlike all the tree-level processes considered hitherto,
which are suppressed by powers of $M_S$.

An interesting feature which emerges in our calculation is the presence of
infrared (IR) divergences (for $d = 2, 3$ and 4) in the
summation over masses of particles in the KK towers. These are due to
the zero modes of the graviton (and radion) KK states and arise
ultimately from the fact that in the IR limit
when the graviton momenta are comparable to
the mass gap between neighbouring excitations, the continuum approximation
is invalid. One is therefore led to believe that the IR
divergences are spurious and should cancel out of a complete 
calculation\footnote{
One should also note that gravity is known to be infrared-safe, especially
in the full (4+$d$)-dimensional theory. Compactification corresponds to
replacing certain components of momentum by masses and this cannot generate 
IR divergences. The authors are grateful to Ashoke Sen for bringing this
fact to their notice.}
A useful suggestion made by the authors of Ref.~\cite{HaLyZh} is to
regulate these expressions by simply dropping the zero mode (whose
coupling is Planck-scale suppressed) and start from the first massive KK
mode.  Cancellation of the IR divergence will therefore manifest itself as
a cancellation of the dependence of the final result on the mass of this
mode. We shall see that with the present level of knowledge, the presence 
of these IR divergences makes the $\rho$-parameter calculation
non-predictive for $d = 2, 3$ and 4. However, these same IR
divergences allow us to make a concrete prediction for $d = 5$ and
6, even without a complete knowledge of the relevant Feynman rules.

The plan of this article is as follows. Section 2 consists of a
micro-review of phenomenological studies of the AADD model carried out
at the tree level. In the next section, we describe the basic formalism 
required
to calculate the $W,Z$ self-energy corrections (and hence the $\rho$
parameter) in the AADD model. Section 4 is devoted to numerical results
from the calculation and their interpretation. It is in this section that
we discuss the cancellation of IR divergences.  We summarize our results
in Section 5. Appendix A comprises of the relevant Feynman rules.  The
machinery of one-loop integrals using the ultraviolet (UV) cutoff $M_S$ is
described in detail in Appendix B. \\

\begin{center}
{\large\bf 2. Phenomenological Studies}
\end{center}

Since gravitons (and radions) couple to any particle with a non-vanishing
energy-momentum tensor, it is possible to make a variety of
phenomenological studies of the interactions of their KK modes. Though
these have not yet been completely explored, several important results are
already available. These investigations can be classified into two types:
those involving {\em real} KK graviton production, and those involving
{\em virtual} graviton exchange. A real KK mode of the graviton will have
individual interactions with matter suppressed by $M^{(4)}_{Pl}$ and will,
therefore, escape any detector of terrestrial proportions. One can,
therefore, see signals~\cite{realgraviton} with large missing momentum and
energy if an observable particle is produced in association with a tower
of KK gravitons.  Cross-sections for these depend explicitly on $d$, the
number of extra dimensions. Each process can be used to obtain a bound on
the string scale $M_S$ for a given $d$.  The most dramatic of these is
$M_S > 50$ TeV for $d = 2$ and it comes from a study~\cite{CuPe} of the
flux of neutrinos from the supernova SN1987A.  However, this astrophysical
bound drops to about a TeV as soon as we go to $d > 3$. Most of the other
processes lead to lower bounds of about 1--1.1 TeV on the string scale for
$d = 2$, and are even weaker for $d > 3$.

Virtual (KK) graviton exchanges lead to extra contributions to processes
involving SM particles in the final state and can be observed as
deviations in the cross-sections and distributions.  After summing over
all the KK modes of the graviton, a tree-level Feynman amplitude involving
graviton exchange is proportional to $\lambda/M^4_S$, where
\begin{eqnarray}
|\lambda| & \simeq & \log(M_S^2/s) \ \ \ {\rm for}~d = 2 \ , \\
          & \simeq & (d -2)^{-1}  \ \ \ \ \  {\rm for}~d > 2 \ ,
\nonumber \label{c4}
\end{eqnarray}
to the leading order in $s/M_S^2$, where $\sqrt{s}$ is the scale of the
interaction~\cite{GiRaWe,HaLyZh}.  It has been found convenient to absorb
the magnitude of $\lambda$ into $M_S^{-4}$, so that bounds from virtual
processes can be presented in a compact form. However, it should be borne
in mind that the effective string scale $M_S$ is, then, $\widetilde M_S =
\lambda^{-1/4} M_S$, and hence different from the actual $M_S$, such as is
considered for real graviton emission. Once this slight abuse of notation
is understood, each process can be used~\cite{virtualgraviton} to obtain a
bound on $\widetilde M_S$ for a given sign of $\lambda$. Among the most
stringent of these bounds are $\widetilde M_S > 1$ TeV for $\lambda = \pm 1$
which come from studies of experimental data on dileptons~\cite{GMR} and
dijets~\cite{MRS3} at the Fermilab Tevatron.

A summary of the most stringent lower bounds from collider data on $M_S$,
as well as the astrophysical bound from a study of the neutrino flux from
the supernova SN1987A, is given in Table 1.

\footnotesize
$$
\begin{array}{|l|c|c|c|c|c|c|} \hline
{\rm Process} 
& \widetilde M_S               & d = 2     & d = 3 & d = 4 & d = 5 & d = 6 \\
\hline
p\bar p \to \ell^+\ell^- & 
  \sim 1.0               & \sim 1.2  & 1.0   & 0.84  & 0.76  & 0.71   \\
\hline
p\bar p \to {\rm dijets} & 
\sim 1.1                 & \sim 1.3  & 1.1   & 0.92  & 0.84  & 0.78   \\
\hline
{\rm astrophysical~bound} &          & 
\sim 50                   & \sim 4   &\sim 1 &       &                \\
\hline
\end{array}
$$
\noindent
\noindent {\bf Table 1}.
{\footnotesize\it Illustrating lower bounds on the string scale $M_S$ (in
TeV) from dilepton~\cite{GMR} and dijet~\cite{MRS3} data at the Fermilab
Tevatron.  The astrophysical bounds from a consideration~\cite{CuPe} of
the neutrino flux at the supernova SN1987A are also shown for purposes of
comparison.  }
\vskip 20pt
\normalsize

\begin{center}
{\large\bf 4. Calculation of Self-Energies}
\end{center}

In electroweak theory, the $\rho$ parameter is defined by~\cite{rho}
\begin{equation}
\rho = \frac{M_W^2}{M_Z^2 \cos^2\theta_W}
 \label{rho_def}
\end{equation}
to all orders in perturbation theory. The most convenient formalism for
the study of this parameter is that given by Peskin and
Takeuchi~\cite{PesTak}, who define $\rho - 1 = {\alpha}~T$, 
where $\alpha$ is the fine structure constant and $\rho$ is
calculated in the zero momentum limit. The $T$-parameter defined
above is one of the so-called {\em oblique} parameters and is zero in
the Standard Model at tree-level. Non-vanishing values of $T$
measure, therefore, one-loop corrections to the $W,Z$ masses in the
SM and/or any new physics effects in these masses. The current
experimental bounds on the (dimensionless) $T$-parameter are
$ T = - 0.21 \pm 0.16 ~(+0.10)$.  

In terms of the self-energy corrections $\Pi(p^2)$ of the gauge
bosons $W^\pm$ and $Z^0$, the $T$-parameter is~\cite{PesTak}
\begin{equation}
T = {\frac {4\pi} {M_W^2} }~\bigg[ \Pi_{WW}(0) 
- \Pi_{ZZ}(0) \cos^2{\theta_W} \bigg]
 \label{T_def}
\end{equation}
where $p$ is the propagator momentum and the self-energies are calculated in
the limit $p^2 \to 0$. This will have small contributions in the Standard 
Model. The excess contribution to the $T$-parameter in
the AADD model (henceforth denoted simply $\delta T$) can be
calculated simply by evaluating the self-energy corrections of the $W$ and
$Z$ bosons due to graviton (radion) loops.

\begin{figure}[htb]
\begin{center}
\vspace*{0.5in}
      \relax\noindent\hskip -4.4in\relax{\includegraphics{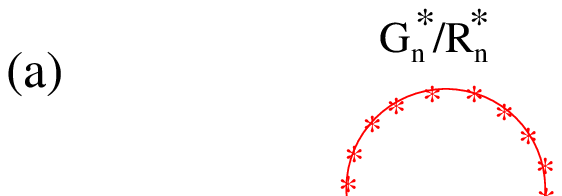}}
\end{center}
\end{figure}
\vspace*{3.2in}
\noindent {\bf Figure 1}.
{\footnotesize\it Feynman diagrams corresponding to self-energy
corrections of the $W$-boson in the AADD model at the one-loop level.
$G_n$ and $R_n$ correspond to graviton excitations with spin 2
and 0 respectively. There will be a similar set of diagrams for the 
$Z$-boson.} %
\vskip 5pt

The Feynman diagrams corresponding to the self-energy of the $W$-boson are
shown in Figure 1($a$-$d$). Since the low energy theory is just linearized
Einstein gravity, we expect it to be non-renormalizable. Hence, rather
than work with a whole family  of effective operators \footnote{such as,
for example $\bar f f WW$-type and $\bar f f ZZ$-type terms; there are
many other possibilities.}, we work with
the unrenormalized vertices as given in Ref. \cite{HaLyZh} and it is
therefore necessary to include tadpole graphs such as those of Figure
1($b$) and 1($c$). These can have all kinds of particles in the loops,
including ($b$) all SM particles and ($c$) the quanta of linearised
gravity.  Moreover (as pointed out in Ref. \cite{HaLyZh}) it is necessary
to include `seagull'-type diagrams such as that in Figure
1($d$). It is important to note that while the diagrams in Figure
1($a$-$c$) involve ${\cal O}(\kappa)$ vertices, the diagrams in Figure
1($d$) involve ${\cal O}(\kappa^2)$ vertices. Unfortunately, Feynman rules
are not readily available in the AADD model for ($i$) the self-couplings in
the pure gravity sector, which are required for the evaluation of
Figure 1($c$) and ($ii$) the full set of ${\cal O}(\kappa^2)$ vertices.
Thus, any calculation of the self-energies at this stage must be
incomplete to this extent.  However, as we shall show, it is
still possible to obtain some meaningful results, and this is the theme of
this work.

Evaluation of the diagrams in Figure 1($a$) and ($b$), using the Feynman
rules given in Appendix A, is a long and tedious process. Some of the
formalism is developed in Appendix B. The final result for
$\Pi_{WW}^{AADD}(0)$ takes the form
\begin{equation}
\Pi_{WW}^{AADD}(0)  \simeq  
\frac{M_S^2}{720\pi} ~\frac{1}{(2\sqrt{\pi})^d}
~\frac {1}{\Gamma(d/2)} 
\int_{x_0}^1 dx ~x^{-1 + d/2} \bigg[ J_W^G(x) 
- \frac{d-1}{d+2}~J_W^R(x) \bigg] 
 \label{Pi_WW}
\end{equation}
where $x = M_n^2/M_S^2$. The functions $J_W^{G,R}(x)$, which arise from
evaluation of the loop integrals in the case of graviton ($G$) and radion
($R$) respectively, are defined in Appendix B. In the above formula
$x_0$ is an IR cutoff which is taken to be $x_0 = (R_c M_S)^{-2}$ following
the suggestion of Ref.~\cite{HaLyZh}.  It corresponds to the mass of the
first massive KK state of the graviton/radion. Using the relation
\cite{ADD3} between $R_c$, $M_S$ and $d$, it follows that
\begin{equation}
x_0 \simeq 2^{1/d} \times 10^{-62/d} 
~\left( \frac{M_S}{\rm 1~TeV} \right)^{4/d} \ ,
\end{equation}
which is clearly minute for $M_S \sim 1$ TeV and approaches unity when
$M_S \to M^{(4)}_{Pl}$.

Similarly, the final result for $\Pi_{ZZ}^{AADD}(0)$ takes the form
\begin{equation}
\Pi_{ZZ}^{AADD}(0) 
\simeq \frac{M_S^2}{720\pi} ~\frac{1}{(2\sqrt{\pi})^d}
~\frac {1}{\Gamma(d/2)} 
\int_{x_0}^1 dx ~x^{-1 + d/2} \bigg[ J_Z^G(x) 
- \frac{d-1}{d+2}~J_Z^R(x) \bigg] 
 \label{Pi_ZZ}
\end{equation}
where, as before, the functions $J_Z^{G,R}(x)$ are defined in Appendix
B.  Substitution of Equations (\ref{Pi_WW}) and (\ref{Pi_ZZ}) in Equation
(\ref{T_def}) now yields the value of $\delta T$ for a given $d$
and value of $M_S$. Since the functions $J_{W,Z}^{G,R}(x)$ are
extremely complicated (see Appendix B), the integrals in the above
formulae are evaluated numerically.

Evaluation of the Feynman diagrams in Figure 1($c$) and ($d$) is not
directly possible because the Feynman rules are still not available.
However, we can make an approximate estimate of the `seagull'-type graph
using the Feynman rule given in Eqn. (84) of Ref. \cite{HaLyZh} for a pair
of (spin-2) gravitons coupling to a pair of scalars. Assuming that the
longitudinal components of the $W$ and $Z$ bosons are given by the
Goldstone equivalence theorem, we can evaluate the `seagull' graph. 
This adds to the function $J_{W,Z}^G(x)$ by a term which we
denote $K_{W,Z}^G(x)$. The full form of $K_{W,Z}^G(x)$ is given in
Appendix B. We then make the assumption that the two contributions due
to the transverse modes are each equal to that due to the longitudinal
mode. This corresponds to multiplying $K_{W,Z}^G(x)$ by a factor of 3. We
next parametrize the error due to these approximations by introducing an
unknown parameter $\xi$, so that the extra contribution reads
\begin{equation}
J_{W,Z}^G(x)  ~~\longrightarrow ~~J_{W,Z}^G(x) ~+~ 3~\xi~K_{W,Z}^G(x)
\end{equation}

We next note that the diagrams in Figure 1($c$) have the {\em same}
particles running in the loop as the corresponding ones in Figure 1($d$),
and therefore, should be proportional to the same loop integrals. In fact,
it will not matter which of the graviton states contributes to the loop
integral, since the propagators are same, each having an identical tower
of states. Of course, there will be explicit dependence on the
graviton/radion in the vertex and in the propagator
connecting the loop with the $WW$ vertex. However, to a first
approximation, we assume that all these can be lumped into a common
contribution, also proportional to $K_{W,Z}^G(x)$. All the extra
contributions due to the diagrams in Figure 1($c$) are thus absorbed into
the unknown parameter $\xi$.  Any momentum dependence arising from the
vertices and propagators contributing to $\xi$ can be removed by defining
$\xi$ to be an average value. We therefore, do not add any extra 
contribution to $J_{W,Z}^R(x)$. 

While the above procedure does not sound very rigorous, it is the best
that can be done until all the relevant Feynman rules become available.
Calculation of these is a non-trivial task and it may be some time before
they are available in the literature \cite{DaJaPaRa}. In principle, the
value of $\xi$, which parametrizes our present state of ignorance of the
effective 
theory to ${\cal O}(\kappa^2)$, can depend explicitly on $M_S$ and on $d$. 
However, as we show in
the next section, any such dependence is very weak, and therefore, we
obtain an \'a posteriori justification for the approximation made here.

It is clear from the above that we have little or no \'a priori knowledge
of what the parameter $\xi$ should be, except that it should be of order
unity (which follows from dimensional arguments). However, it is here that
the IR divergences come to our help. We shall see in the following section
that gravitonic contributions to $\delta T$ are IR divergent for $d = 2,
3$ and $4$. It has been argued above that these IR divergences are
spurious and hence we demand that they should cancel out when we sum over
all diagrams. Since $J_{W,Z}^{G,R}(x)$ and $K_{W,Z}^G(x)$ are
individually IR divergent, we can {\em tune} the value of 
$\xi$ to get cancellation of
these divergences. We then claim that this is the value of $\xi$ which
would be obtained from a proper knowledge of the Feynman rules in this
model\footnote{This is in the same spirit as the determination of, for
example, the three-gluon vertex in QCD from considerations of gauge
invariance}. The exact procedure is described in the following section.

\begin{center}
{\large\bf 4. Numerical Results and Discussion}
\end{center}

Numerical evaluation of the integrals in Equations (4) and (6) is not 
completely
straightforward. To see this, consider, for example, $M_S =$ 1~TeV and $d
= 2$, in which case, the IR cutoff comes out to be $x_0 \sim 10^{-31}$.
The integral runs, therefore, over several orders of magnitude and, being
IR divergent, receives its principal contribution from the smallest values
of $x_0$. It is, therefore, convenient to smoothen the integrand by
the transformation $\xi = - \log x$, which allows the integration to be
done using standard numerical techniques. For $d \leq 4$, the IR cutoff
$y_0 = - \log x_0$ grows with $M_S$ (corresponding to the increase in
energy gap between the two lowest KK states), so that, for very large
values of $M_S \sim M^{(4)}_{Pl}$ (which means $x_0 \simeq 1$) the range
of integration gets pinched off. However, the pinch-off is very sharp and 
only
takes place when $x_0 \simeq 1$ to great precision. Of course, such large
values of $M_S$ are uninteresting from the experimental point of view, but
it is reassuring to note that even if we get unacceptably large values for
the $\rho$-parameter at low values of $M_S$, there is always the
possibility of having extra dimensions compactified to the Planck length.

In their leading order calculation of one-loop effects in the AADD
model, the authors of Ref.~\cite{HaLyZh} have shown that these
corrections to the scalar mass can be written in terms of
(dimensionless) loop integrals, of which two are
\begin{equation} 
I_4 = \int_{x_0}^1 dx~x^{-2 + d/2} \ , \qquad 
I_5 = \int_{x_0}^1 dx~x^{-3 + d/2} \ ,
\end{equation} 
where $x_0$ is the IR cutoff described above. On evaluation these
come out to be
\begin{equation}
I_4 = \frac{2}{d-2} \bigg( 1 - x_0^{-1+d/2} \bigg) \ , \qquad 
I_5 = \frac{2}{d-4} \bigg( 1 - x_0^{-2+d/2} \bigg) \ . 
\end{equation} 
Now, these are clearly IR divergent for $d < 2$ and $d < 4$ respectively.
Even the use of the IR cutoff $x_0$ leads to extremely large values of
these integrals, since it is clear that $x_0$ is an extremely small
quantity (unless $M_S \to M^{(4)}_{Pl}$.

We now adopt the philosophy that the infrared divergences are spurious and
must cancel out of a complete calculation. The main motivation for this
--- explained above --- is that gravity is well-known to be infrared-safe,
in fact, more so in dimensions greater than four, and this feature is
expected to be preserved under compactification. The
origin of the IR divergences may be attributed to a breakdown of the
mass-continuum approximation in the low-momentum limit. Since we cannot
make a complete analytic calculation and check the cancellation explicitly,
the best we can do is to {\em tune} the value
of the unknown parameter $\xi$ and see if we can achieve cancellations in
the different cases for $d$ and $M_S$. Our results are shown in Figure~2,
for $M_S = 1$ TeV, $M_H = 250$ GeV and $d = 2, 3$ and 4.

\begin{figure}[htb]
\begin{center}
\vspace*{0.5in}
      \relax\noindent\hskip -3.4in\relax{\includegraphics{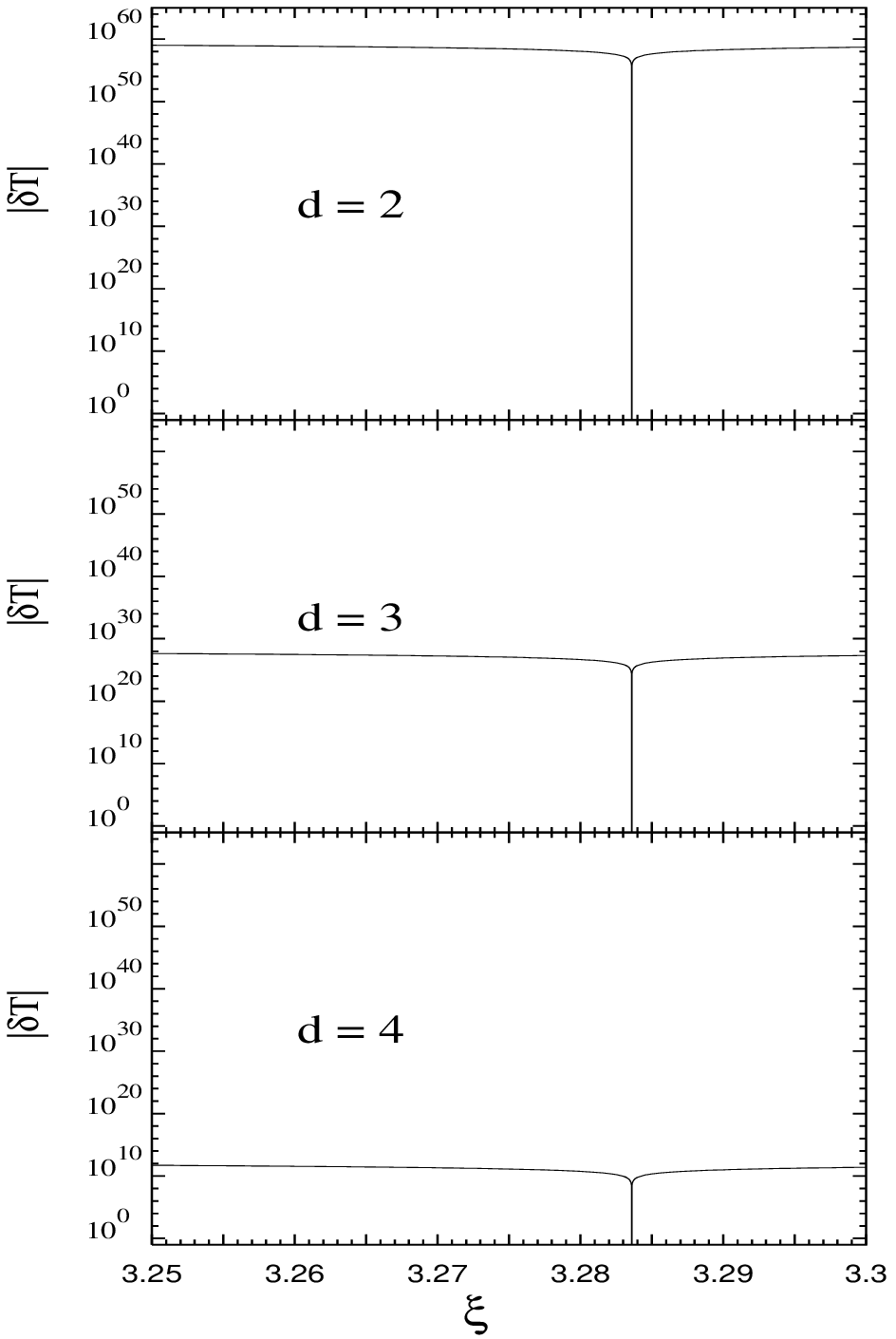}}
\end{center}
\end{figure}
\vspace*{4.6in}
\noindent {\bf Figure 2}.
{\footnotesize\it Illustrating the cancellation of IR divergences in
$\delta T$ for $d = 2, 3$ and 4. Note that these occur for the same
value of $\xi$ in each case. This is also insensitive to variations in
$M_S$ and $M_H$.  }
\vskip 5pt

In Figure 2, it may be seen that a sharp cancellation occurs for a
specific value $\xi \simeq 3.2836$.  Amazingly enough, we get {\em the
same value} for all the three values of $d$. We also find that the value
of $\xi$ for which the cancellation occurs is almost completely
insensitive to the value of $M_S$, while it has a very weak dependence on
the mass of the Higgs boson (0.2\% as $M_H$ varies from 100 GeV to 700
GeV). This constant value of $\xi$ may be attributed to the fact that it
is almost entirely made up of constant or averaged factors in the unknown
coupling, including $\kappa, R_c$ and the gauge boson masses, apart from
purely numerical factors. We believe that a complete calculation, when
available, will predict precisely this value $\xi \simeq 3.2836$. It is
reassuring to note that the value of $\xi$ is indeed of order unity as
we expect. 

Obviously, when such fine cancellations take place, it is not possible to
predict the value of $\delta T$ for $d = 2, 3$ and 4, since any result can
be obtained by making small changes in $\xi$.  However, in view of the fact
that $\xi$ is almost constant for $d = 2, 3$ and 4, we feel emboldened to
extrapolate this value to the (IR-finite) cases $d = 5$ and 6 as well. 
We thus feed in this value of $\xi$ for the cases $d
= 5$ and 6, and are able to make
concrete predictions for $\delta T$ in the AADD model. Figure 3 shows our
results for $\delta T$ as a function of $M_S$ ($a$) in the range of
interest for collider experiments and ($b$) in the region where the 
UV-divergent integral gets pinched-off as $x_0 \to 1$.  As before, we set
$M_H = 250$ GeV, though there is little variation of the results with this
parameter.

\begin{figure}[htb]
\begin{center}
\vspace*{1.0in}
      \relax\noindent\hskip -3.4in\relax{\includegraphics{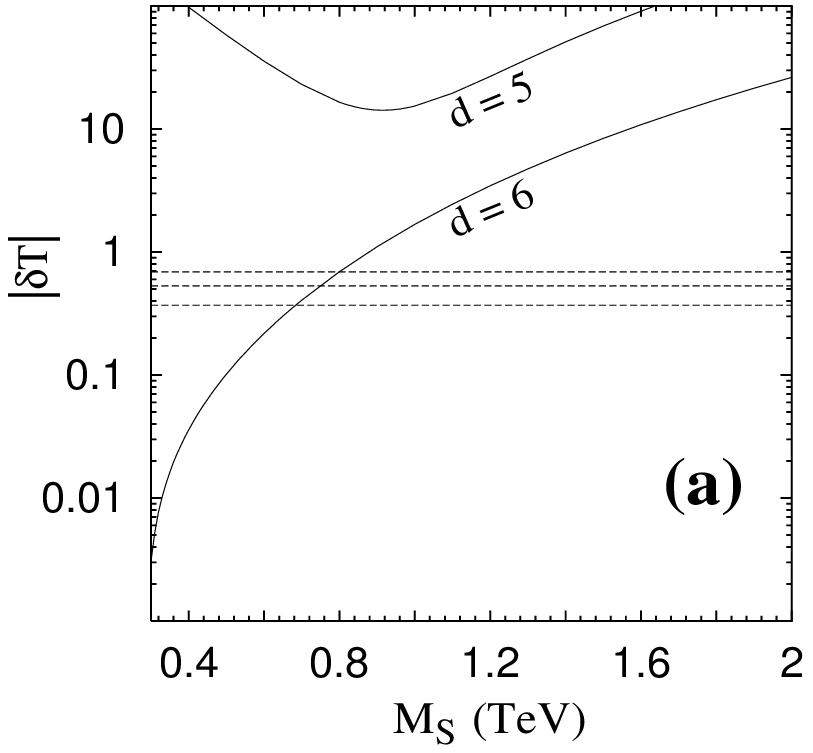}}
      \relax\noindent\hskip  3.4in\relax{\includegraphics{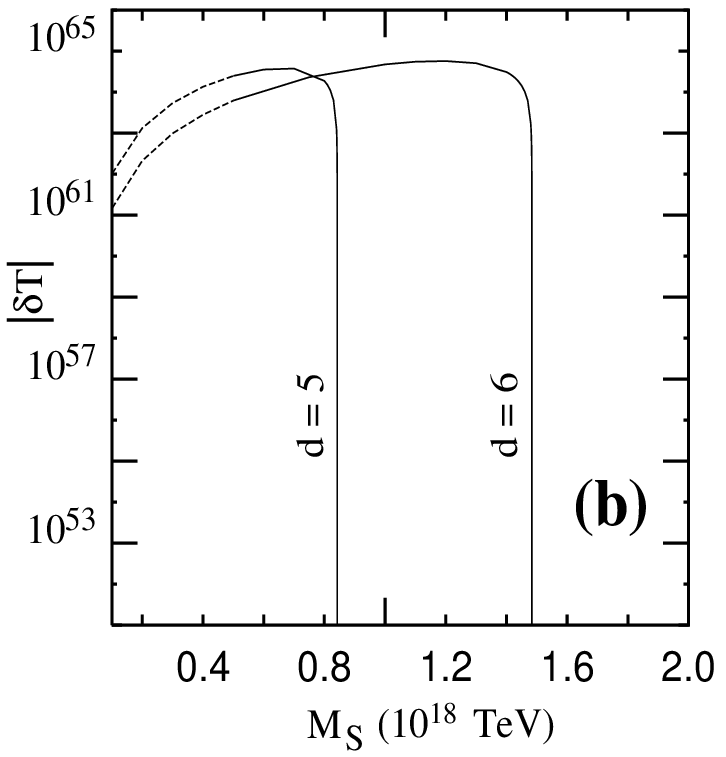}}
\end{center}
\end{figure}
\vspace*{1.5in}
\noindent {\bf Figure 3}.
{\footnotesize\it Variation in $\delta T$ in the AADD model with
increasing $M_S$, setting $\xi = 3.2836$. In ($a$), the dashed lines 
correspond (from bottom upwards) to the experimental upper limits at 
$1\sigma$, $2\sigma$, $3\sigma$ respectively. The pinching-off of the
UV-divergent integral at the Planck scale is illustrated in ($b$).}
\vskip 5pt

As Figure 3 shows, the gravitational contribution to the $T$-parameter
shows a non-decoupling behaviour, though, of course, it finally falls to
zero when $M_S \to M^{(4)}_{Pl}$. The non-decoupling behaviour, we feel,
is an indication that the theory is an effective one. In fact, for $d =
5$, it appears that any value of $M_S$ in the range displayed in 
Figure 3($a$) is ruled out
by experiment --- which forces us to have $M_S \sim M^{(4)}_{Pl}$, as in 
Figure 3($b$).  
For $d = 6$ we obtain an {\em upper} bound of about 725 GeV at 95\% C.L., 
which is
barely allowed by experimental data (see Table 1). A slight improvement in
the data is likely to rule this out and this would force $M_S \sim
M^{(4)}_{Pl}$ for $ d = 6$ as well. 

The actual scenario need not be so gloomy, however, since the effective
theory of Refs.~\cite{GiRaWe} and \cite{HaLyZh} is only an approximation
and the underlying (string?) theory is believed to be finite.  At the
present juncture, it is not possible to make calculations in the exact
theory, and it would, therefore, be necessary to construct a
phenomenological extension of the effective theory, by adding extra
(non-renormalizable) operators with unknown coefficients\footnote{This
would be analogous to the twelve ${\cal O}(p^6)$ operators in chiral
perturbation theory.} to the Lagrangian~\cite{nonrenorm}. The coefficients
may then be adjusted to produce acceptable values of the $\rho$ (and $T$)
parameter, even for $d = 5$ and 6. This is a messy business which smacks
of fine-tuning and will not be attempted in this work.

Finally, it is worth noting that in all these calculations, the
contribution due to the radion, which has much lower levels of
divergence, both in the UV and IR regions, turns out to be negligible
in comparison to that of the graviton. Its inclusion in all the above
formulae is more for the sake of completeness that for numerical
accuracy.

\begin{center}
{\large\bf 6. Conclusions}
\end{center}

In this work, we have performed a careful study of the $W$ and $Z$
boson self-energy corrections using the effective theory of KK graviton
(and radion) interactions developed in Ref.~\cite{HaLyZh} from the
original suggestion of Refs.~\cite{ADD1,ADD2}.  Since these are one-loop
calculations, the results have several interesting features, which do not
show up in tree-level calculations. The results turn out to be strongly
divergent, both in the UV and IR regimes. Using a cutoff at either end, as
suggested in Ref.~\cite{HaLyZh}, can lead to large values of the
$T$-parameter (and hence of the $\rho$ parameter). Knowing gravity to be
IR-safe, we demand cancellation of the IR divergences and use this to
determine the unknown $\xi$ which parametrizes our ignorance of the full
set of Feynman rules in this model. Interestingly, we find that $\xi$ is 
independent of
$M_S$ and the number of extra dimensions, and hence can be used to make
concrete predictions for 5 and 6 extra dimensions. We then find that for
$d = 5$, $M_S$ is driven by the $T$-parameter constraint to the Planck
scale, while for $d = 6$, a small window in $M_S$ around 725 -- 750 GeV is
still viable.  We are unable to make any predictions for $ d < 5$.

We wish to emphasize that while our results do seem to show that models
with large extra dimensions do not work well for $d > 4$, all that this
means is that the formalism of Ref.~\cite{HaLyZh} does not work well for
$d > 4$.  A deeper (perhaps finite) theory should certainly provide better
insights and more acceptable values of the $\rho$-parameter. It might also
be possible to build a phenomenological theory by adding
non-renormalizable operators to the effective Lagrangian, but that lies
outside the scope of this work.

\noindent
{\em Note added:} After this work was completed, we received the
preprint in Ref. \cite{Zhang} where a similar calculation has been
done, using a purely analytic approach. The rather striking
difference in our results from theirs is probably due to the fact
that tadpole diagrams, as in Fig. 1 ($b$) and ($c$) have not been
taken into account in Ref. \cite{Zhang}. We thus believe that the 
calculation
in Ref. \cite{Zhang}, though elegant, is incomplete and should not be
taken as evidence for a decoupling behaviour in $M_S$, at least for the
$\rho$ parameter.

\begin{center}
{\bf Acknowledgements}
\end{center}

The authors would like to thank Kingman Cheung, Debajyoti Choudhury, Dilip
Kumar Ghosh, Sourendu Gupta, Pankaj Jain, Uma Mahanta, Sudipta Mukherji,
Sukanta Panda, Gautam Sengupta and Ren-Jie Zhang for useful discussions.

\newpage
\begin{center}
{\large\bf Appendix A: Feynman rules}
\end{center}

\setcounter{equation}{0}
\renewcommand{\theequation}{A.\arabic{equation}}

The Feynman rules relevant for the calculation of the $\rho$
parameter are as follows. The vertex for a pair of gauge bosons
coupling to the graviton are given by an effective Lagrangian
\begin{equation} 
{\cal L}_{eff} = 
i \bigg( T_{GW}^{\mu\nu\rho\sigma} G^{(n)}_{\mu\nu} 
~+~ T_{RW}^{\rho\sigma} \Phi^{(n)}
\bigg)
\bigg[ W^+_\rho W^-_\sigma + \frac{1}{2}~Z_\rho Z_\sigma 
+ \frac{1}{2}~A_\rho A_\sigma \bigg] \ .
\end{equation}  
The graviton coupling $T_{GW}^{\mu\nu\rho\sigma}(p,q)$ is given by
\begin{equation}
T_{GW}^{\mu\nu\rho\sigma}(p,q) 
= - \kappa \bigg[ (M_{W/Z}^2 + p.q) C^{\mu\nu\rho\sigma} 
+ D^{\mu\nu\rho\sigma}(p,q) \bigg] \ ,
\end{equation}
where $\kappa = \sqrt{16\pi G_N}$ and
\begin{eqnarray}
C^{\mu\nu\rho\sigma} & = & 
\eta^{\mu\rho} \eta^{\nu\sigma} + \eta^{\mu\sigma}  \eta^{\nu\rho}
- \eta^{\mu\nu} \eta^{\rho\sigma} \ , \\
 D^{\mu\nu\rho\sigma}(p,q) & = &
\eta^{\mu\nu} p^\sigma q^\rho  \nonumber \\
&& - \bigg[ (\eta^{\mu\sigma} p^\nu q^\rho + \eta^{\mu\rho} p^\sigma q^\nu
   - \eta^{\rho\sigma} p^\mu q^\nu ) + (\mu \leftrightarrow \nu) \bigg] \ .
\end{eqnarray}
In the above equation, both $W/Z$ momenta are assumed to be directed
towards the vertex. The radion coupling is given by
\begin{equation}
T_{RW}^{\rho\sigma} = \sqrt{\frac{8}{3(2 + d)}}~~\kappa M_{W/Z}^2 
~~\eta^{\rho\sigma} \ .
\end{equation}

The propagator for a graviton with momentum $k$ is 
\begin{equation}
\frac{i P_{\mu\nu\rho\sigma}(k)}{k^2 - M_n^2 + i\epsilon}
\end{equation}
where the polarisation sum $P_{\mu\nu\rho\sigma}$ is given by
\begin{eqnarray}
P_{\mu\nu\rho\sigma}(k) = 
& + & \frac{1}{2} \bigg( \eta_{\mu\rho} - \frac{k_\mu k_\rho}{M_n^2} \bigg)
            \bigg( \eta_{\nu\sigma} - \frac{k_\nu k_\sigma}{M_n^2} \bigg)
\nonumber \\
& + & \frac{1}{2} \bigg( \eta_{\mu\sigma} - \frac{k_\mu k_\sigma}{M_n^2} \bigg)
              \bigg( \eta_{\nu\rho} - \frac{k_\nu k_\rho}{M_n^2} \bigg) 
\nonumber \\
& - &\frac{1}{3} \bigg( \eta_{\mu\nu} - \frac{k_\mu k_\nu}{M_n^2} \bigg) 
            \bigg( \eta_{\rho\sigma} - \frac{k_\rho k_\sigma}{M_n^2} \bigg)
\end{eqnarray}
Finally, the propagator for a radion is given by
\begin{equation}
\frac{i (d - 1)}{k^2 - M_n^2 + i\epsilon} \ ,
\end{equation}
where the extra factor $(d - 1)$ arises from the sum over different
modes of the radion in each dimension. 

In this Appendix we have closely followed the notation and conventions of 
Ref. \cite{HaLyZh}. 

\begin{center}
{\large\bf Appendix B: Loop Integrals}
\end{center}
\setcounter{equation}{0}
\renewcommand{\theequation}{B.\arabic{equation}}

The momentum integrals arising in one-loop calculations involving KK
graviton and radion modes are calculated using an ultra-violet cutoff
$M_S$, where $M_S$ is the `string' scale. We use the well-known notation
of 'tHooft and Veltman and Passarino and Veltman~\cite{Veltman} to
describe momentum integrals of products of propagator functions. Then, the
scalar one-point function turns out to be
\begin{eqnarray} 
A(m^2) & = & \int \frac {d^4 k}{\pi^2} ~\frac {1}{k^2 + m^2 + i \epsilon} 
\nonumber \\
& = & M_S^2 ~\widetilde{A}\biggl(\frac {m^2}{M_S^2}\biggr)
\end{eqnarray}
where 
\begin{equation}
\widetilde{A}(x) = 1 - {x}~\log (1 + \frac {1}{x})
\end{equation}
As in Refs.~\cite{Veltman}, all integrals are defined in Euclidean
space.

For the calculation of $T$, we require to calculate the boson
propagators in the zero momentum limit. Thus, for the scalar
two-point function, we define
\begin{eqnarray}
\widetilde{B}_0(x_1,x_2) = B_0(m_1^2, m_2^2;0) 
& = & \lim_{p^2 \rightarrow 0} \int {\frac {d^4 k}{\pi^2}} 
~{\frac {1}{[k^2 + m_1^2]~[(k + p)^2 + m_2^2]}}  \nonumber \\
& = &  \frac {-1}{(x_1 - x_2)} 
\bigg[\widetilde {A}(x_1) - \widetilde {A}(x_2)\bigg]
\end{eqnarray}
where $x_{1,2} = m_{1,2}^2/M_S^2$. We are then in a position
to define
\begin{eqnarray}
B_0^\prime (m_1^2,m_2^2; 0) & = &
\lim_{p^2 \to 0} \frac{\partial}{\partial p^2} B_0(m_1^2,m_2^2; p^2) 
\nonumber \\
& = & \frac{1}{M_S^2} ~\widetilde B_0^\prime (x_1,x_2)
\end{eqnarray}
which turns out, on evaluation, to be 
\begin{eqnarray}
\widetilde {B}_0^\prime(x_1,x_2) & = & 
\frac {1}{x_1 x_2 (x_1 - x_2)^3}  \\
& \times & ~\bigg[x_1 - x_2 - x_1 x_2 \log \biggl(\frac {x_1}{x_2}\biggr) 
- (1 + x_1 x_2)~\{x_1 \widetilde{A}(x_2) - x_2 \widetilde{A}(x_1) \}
\bigg] \ . \nonumber 
\end{eqnarray}

The vector two-point function is defined to be 
\begin{eqnarray}
B_\mu(m_1^2,m_2^2;p) & = & 
\int \frac {d^4 k}{\pi^2}~\frac{k_\mu}{[k^2 + m_1^2]~[(k + p)^2 + m_2^2]} 
\nonumber \\
& = & p_\mu~B_{1}(m_1^2,m_2^2;p^2) \ ,
\end{eqnarray}
which leads to 
\begin{eqnarray} 
\widetilde {B}_1(x_1,x_2) & = & B_1(m_1^2,m_2^2;0) \nonumber \\
& = & -~\frac{1}{2}\bigg[ \widetilde {B}_0(x_1,x_2) 
~+~ (x_1 - x_2)~\widetilde {B}_0^\prime(x_1,x_2)\bigg] \ .
\end{eqnarray}

The tensor two-point function is
\begin{eqnarray}
B_{\mu\nu}(m_1^2,m_2^2;p) & = & 
\int \frac {d^4 k}{\pi^2}~\frac{k_\mu k_\nu}{[k^2 + m_1^2]~[(k +p)^2 + m_2^2]} 
 \\
& = & p_\mu p_\nu~B_{21}(m_1^2,m_2^2; p^2) 
~+~ \delta_{\mu \nu}~B_{22}(m_1^2,m_2^2; p^2) \ , \nonumber 
\end{eqnarray}
and we define
\begin{equation}
B_{22}(m_1^2,m_2^2; 0) = M_S^2~\widetilde{B}_{22}(x_1,x_2) \ .
\end{equation}

This function can be evaluated as
\begin{equation}
\widetilde{B}_{22}(x_1,x_2) ~=~ \frac{1}{6} 
\bigg[ \widetilde{A}(x_1) ~-~ 2 x_2~\widetilde{B}_0(x_1,x_2) 
~+~ (x_1 - x_2)~\widetilde {B}_1(x_1,x_2) \bigg]
\end{equation}
A similar expression can be derived for $B_{21}$, but it is not
relevant for the present calculation.

Now, in terms of these functions, we can write the integrands relevant for
$\Pi_{WW}(0)$ as 
\begin{eqnarray} 
J_G^W(x) & = &  I_G(X_W,x) \\
& - & \frac{3X_W}{x} 
\bigg\{~283 - 12~X_W \widetilde A(X_W) - 6~X_Z \widetilde A(X_Z) 
- 4~X_H \widetilde A(X_H) \nonumber \\
& & \hspace*{0.9in} +~8~\sum_f~C_f~X_f \widetilde A(X_f) \bigg\} \nonumber
\end{eqnarray}
where $X_i = Mi^2 / M_S^2$ and the colour factor $C_f = 3$ for quarks
and 1 for leptons. Similarly
\begin{eqnarray} 
J_D^W(x) & = &  4 X_W \bigg[ 16~I_D(X_W,x) \\
&  & \hspace*{0.3in} - \frac{6}{x}
\bigg\{~295 + 24~X_W \widetilde A(X_W) + 6~X_Z \widetilde A(X_Z) 
+ 2~X_H \widetilde A(X_H) \nonumber \\
& & \hspace*{0.9in} +~8~\sum_f~C_f~X_f \widetilde A(X_f) \bigg\} \bigg] 
\nonumber
\end{eqnarray}

Similarly, we can write the integrands relevant for $\Pi_{ZZ}(0)$ as 
\begin{eqnarray} 
J_G^Z(x) & = &  \frac{1}{4}I_G(X_W,x) \\
& - & \frac{3X_Z}{2x} 
\bigg\{~283 - 12~X_W \widetilde A(X_W) - 6~X_Z \widetilde A(X_Z) 
- 4~X_H \widetilde A(X_H) \nonumber \\
& & \hspace*{0.9in} +~8~\sum_f~C_f~X_f \widetilde A(X_f) \bigg\} \nonumber \\
J_D^Z(x) & = &  4 X_Z \bigg[ 4~I_D(X_W,x) \\
&  & \hspace*{0.3in} - \frac{3}{x}
\bigg\{~295 + 24~X_W \widetilde A(X_W) + 6~X_Z \widetilde A(X_Z) 
+ 2~X_H \widetilde A(X_H) \nonumber \\
& & \hspace*{0.9in} +~8~\sum_f~C_f~X_f \widetilde A(X_f) \bigg\} \bigg] \nonumber
\end{eqnarray}

The $I$-functions, which arise from evaluation of the diagram in
Fig.~1($a$), can be written as
\begin{equation}
I_D(x_1,x_2) 
= {\frac {1}{2}~\widetilde {A}(x_1)} 
+ (3 x_1 - x_2)~\widetilde{B}_0(x_1,x_2) 
+ \frac{1}{2}(x_1 - x_2)~\widetilde {B}_1(x_1,x_2) 
\end{equation}
for the loop integral involving radion exchange, and
\begin{eqnarray}
I_G(x_1,x_2) & = & \frac{1}{x_1 - x_2}~\frac{1}{x_2^2} 
~\bigg[ f_1(x_1,x_2) ~+~ f_2(x_1,x_2)~\widetilde{A}(x_1) 
       ~+~ f_3(x_1,x_2)~\widetilde{A}(x_2) \nonumber \\
   & & \hspace*{1in} ~+~ f_4(x_1,x_2)~\widetilde{B}_{22}(x_1,x_2) \bigg] 
\nonumber \\
\end{eqnarray}
for the loop integral involving graviton exchange. The functions
$f_i(x_1,x_2)$ are given by
\begin{eqnarray}
f_1(x_1,x_2)  & = &  
 (x_1 - x_2)~\bigg[ 12 x_2^3 - 60 x_1 x_2^2 
+ 4 (24x_1^2 - 19 x_1 - 11) x_2 \nonumber \\ 
&& \hspace*{2.4cm} - (48 x_1^3 + 98 x_1^2 + 64 x_1 + 33) \bigg]
\\
f_2(x_1,x_2) & = & 48 x_1 x_2~\bigg[12 x_2^2 ~+~ x_1 x_2 ~-~ 3 x_1^2\bigg]
\\
f_3(x_1,x_2) & = & 12~\bigg[x_2^2 ~-~ 47 x_1^2 x_2^2 ~+~ 10 x_1^3 x_2 
~-~ 4 x_1^4\bigg]
\\
f_4(x_1,x_2) & = & - 48 x_1 (x_1 - x_2)~\bigg[x_2^2 ~+~ 8 x_1 x_2 
~+~ x_1^2\bigg]
\end{eqnarray}

Finally, the `evaluation' of seagull diagrams leads to the functions
\begin{eqnarray}
K^G_W(x) & = & 28.9 ~\frac{X_W}{x} ~\left[ 2 + 21 x 
+ 54 x^2 \widetilde A(x) \right] \ ,
\nonumber \\
K^G_Z(x) & = & 72.0 ~\frac{X_Z}{x} ~\left[ 2 + 21 x 
+ 54 x^2 \widetilde A(x) \right] \ .
\end{eqnarray}


\begin{thebibliography}{99}
\footnotesize

\def\pr#1,#2,#3 { {\em Phys.~Rev.}        ~{\bf #1},  #2 (#3) }
\def\prd#1,#2,#3{ {\em Phys.~Rev.}        ~{\bf D#1}, #2 (#3) }
\def\prl#1,#2,#3{ {\em Phys.~Rev.~Lett.}  ~{\bf #1},  #2 (#3) }
\def\plb#1,#2,#3{ {\em Phys.~Lett.}       ~{\bf B#1}, #2 (#3) }
\def\npb#1,#2,#3{ {\em Nucl.~Phys.}       ~{\bf B#1}, #2 (#3) }
\def\prp#1,#2,#3{ {\em Phys.~Rept.}       ~{\bf #1},  #2 (#3) }
\def\zpc#1,#2,#3{ {\em Z.~Phys.}          ~{\bf C#1}, #2 (#3) }
\def\epj#1,#2,#3{ {\em Eur.~Phys.~J.}     ~{\bf C#1}, #2 (#3) }
\def\mpl#1,#2,#3{ {\em Mod.~Phys.~Lett.}  ~{\bf A#1}, #2 (#3) }
\def\ijmp#1,#2,#3{{\em Int.~J.~Mod.~Phys.}~{\bf A#1}, #2 (#3) }
\def\ptp#1,#2,#3{ {\em Prog.~Theor.~Phys.}~{\bf #1},  #2 (#3) }


\bibitem{ADD1}
N.~Arkani-Hamed, S.~Dimopoulos and G.~Dvali, \plb429,263,{1998}. 
Some precursors of the model are
V.~Rubakov and M.~Shaposhnikov, \plb125,136,{1984};
A.~Barnaveli and O.~Kancheli, {\it Sov. J. Nucl. Phys.} {\bf 51}, 573
(1990); 
I.~Antoniadis, \plb246,377,{1990};
I.~Antoniadis, C.~Mu\~noz and M.~Quiros, \npb397, 515,{1993};
I.~Antoniadis, K.~Benakli and M.~Quiros, \plb331,313,{1994}.

\bibitem{GrScWi}
M.B.~Green, J.~Schwarz and E.~Witten, {\it Superstring Theory}, Vols.
1 and 2 (Cambridge University Press, 1987).

\bibitem{Polchinski}
For an excellent pedagogical review, see, for example,
J.~Polchinski, {\it Tasi Lectures on D-Branes},
hep-th/9611050 (1996).

\bibitem{ADD2}
I.~Antoniadis, N.~Arkani-Hamed, S.~Dimopoulos and G.~Dvali, \plb463,257,{1998}. 

\bibitem{LoChPr} J.C.~Long, H.W.~Chan and J.C.~Price, \npb539,23,{1999}. 

\bibitem{ADD3}
N.~Arkani-Hamed, S.~Dimopoulos and G.~Dvali, \prd59,086004,{1999}.

\bibitem{ArDiMaRu}
N.~Arkani-Hamed, S.~Dimopoulos and J.~March-Russell, SLAC preprint
SLAC-PUB-7949 (1998), hep-th/9809124.

\bibitem{KK} 
G.~Nordstr\"om, {\it Phys. Zeitschr.} {\bf 15}, 504 (1914); 
Th.~Kaluza, {\it Situngober. Preuss. Akad. Wiss. Berlin}, p.966 (1921);
O.~Klein, {\it Z.~Phys.} {\bf 37}, 895 (1926), 1938 Conference on New
Theories in Physics, Kazimierz, Poland (1938) and {\it Helv. Phys. Acta},
Supp.{\bf IV}, 58 (1956); 
For a fascinating discussion of the early
history of gauge and KK theories, see L.~O'Raifeartaigh and N.~Straumann, 
hep-ph/9810524; 
A more technical review is given by D.~Bailin and A.~Love,
{\it Rept. Prog. Phys.} {\bf 50}, 1087 (1987).


\bibitem{GiRaWe}
G.F.~Giudice, R.~Rattazzi and J.D.~Wells, \npb544,3,{1998}. 

\bibitem{HaLyZh}
T.~Han, J.D.~Lykken and R.-J.~Zhang, \prd59,105006,{1999}.

\bibitem{Sundrum}
R.~Sundrum, \prd59,085009,{1999}, \prd59,085010,{1999}.

\bibitem{Graesser} M.~Graesser, LBL preprint LBNL-42812 (Feb 1999),
hep-ph/9902310. 

\bibitem{rho} M. Veltman, {\it Act. Phys. Pol.} {\bf B8}, 475 (1977).

\bibitem{precision} T.G.~Rizzo and J.D.~Wells, SLAC preprint 
SLAC-PUB-8119 (Jun 1999), hep-ph/9906234;
A.~Strumia, Pisa U. preprint IFUP-TH-29-99 (Jun 1999), hep-ph/9906266;
R.~Casalbuoni {\it et al.}, U. of Florence preprint DFF-341/7/99 (Jul 1999),
hep-ph/9907355. 

\bibitem{realgraviton}
E.A.~Mirabelli, M.~Perelstein and M.E.~Peskin, \prl82,2236,{1999}; 
K.~Cheung and W.-Y.~Keung, Davis preprint UCD-HEP-99-6 (1999),
hep-ph/9903294; 
K.~Cheung, Davis preprint UCD-HEP-99-8 (1999), hep-ph/9904266; 
D.~Atwood, S.~Bar-Shalom and A.~Soni, Brookhaven preprint
BNL-HET-99/11, hep-ph/9903538; 
S.~Cullen and M.~Perelstein, \prl83,268,{1999}; 
T.~Han, D.~Rainwater and D.~Zeppenfeld, Madison MADPH-99-1115 (May 1999),
hep-ph/9905423; 
D.K.~Ghosh, P.~Poulose and K.~Sridhar, TIFR preprint TIFR-TH-99-47 (1999).

\bibitem{CuPe}
S.~Cullen and M.~Perelstein in Ref.~\cite{realgraviton},
hep-ph/9903422; 

\bibitem{virtualgraviton}
T.G.~Rizzo, \prd59,115010,{1999};
X.-G.~He, Taiwan preprint (1999), hep-ph/9905295; 
K.~Agashe and N.G.~Deshpande, \plb456,60,{1999};
T.G.~Rizzo, SLAC preprint SLAC-PUB-8114 (1999), hep-ph/9904380;
J.L.~Hewett, \prl82,4765,{1999}; 
P.~Mathews, S.~Raychaudhuri and K.~Sridhar, TIFR preprint
TIFR-TH/99-13 (1999), hep-ph/9904232; 
P.~Mathews, S.~Raychaudhuri and K.~Sridhar, \plb450,343,{1999}; 
T.G.~Rizzo, SLAC preprint SLAC-PUB-7986 (1999), hep-ph/9902273;
K.Y.~Lee, {\it et al.},  Seoul preprint SNUTP-99-022 (1999),
hep-ph/9905227.
P.~Mathews, S.~Raychaudhuri and K.~Sridhar, \plb455,115,{1999};
C.~Balasz, {\it et al.}, Michigan State U. preprint MSUHEP-90105
(1999), hep-ph/9904220; 
K.Y.~Lee, H.S.~Song and J.H.~Song, Seoul preprint SNUTP-99-021 (1999),
hep-ph/9904355;  
T.G.~Rizzo, SLAC preprint SLAC-PUB-8071 (1999), hep-ph/9903475; 
D.~Atwood, S.~Bar-Shalom and A.~Soni, Riverside preprint 
UCRHEP-T258 (Jun 1999), hep-ph/9906400;
X.-G.~He, hep-ph/9905295
P.~Mathews, P.~Poulose and K.~Sridhar, \plb461,196,{1999};
D.K.~Ghosh, P.~Mathews, P.~Poulose and K.~Sridhar, {\it JHEP} {\bf 9911}, 
004 (1999).

\bibitem{GMR} A. Gupta, N.K. Mondal and S. Raychaudhuri, TIFR
preprint TIFR-HECR-99-02 (Apr 1999), hep-ph/9904234. 

\bibitem{MRS3} P.~Mathews, S.~Raychaudhuri and K.~Sridhar 
in Ref.~\cite{virtualgraviton}, hep-ph/9904232. . 

\bibitem{PesTak} M.~Peskin and T.~Takeuchi, \prl65,964,{1990} and 
\prd46,381,{1992}. 

\bibitem{DaJaPaRa} P.~Das, P.~Jain, S.~Panda and S.~Raychaudhuri, {\em
work in progress}.

\bibitem{nonrenorm} K.~Huitu and T.~Kobayashi, Helsinki preprint
HIP-1999-40-TH (Jun 1999), hep-ph/9906431.

\bibitem{PDG} Review of Particle Physics, \epj3,1,{1998}. 

\bibitem{Veltman} G.~'tHooft and M.~Veltman, \npb153,365,{1979};
G.~Passarino and M.~Veltman, \npb160,151,{1979}. 

\bibitem{Zhang} J.R.~Espinosa and R.-J.~Zhang, Madrid preprint 
FT-UAM-CSIC-00-09 (March 2000), hep-ph/0003246. 

\end{thebibliography}
\end{document}